\documentclass[aps,prb,showpacs,twocolumn]{revtex4}
\usepackage{graphicx}
\begin{document}
\title{Microwave modulation of electron heating and Shubnikov-de Haas oscillation
 in two-dimensional electron systems}
\author{X.L. Lei and S.Y. Liu}
\affiliation{Department of Physics, Shanghai Jiaotong University,
1954 Huashan Road, Shanghai 200030, China}

\begin{abstract}
Recently discovered modulations of Shubnikov-de Haas oscillations in microwave-irradiated
 two-dimensional electron systems are shown to arise from electron heating 
induced by the radiation. The electron temperature, obtained by balancing the 
energy absorption from the microwave field and the energy dissipation to the lattice 
through realistic electron-phonon couplings, exhibits resonance. The modulation 
of the Shubnikov de Haas oscillation and the suppression of magnetoresistance are 
demonstrated together with microwave-induced 
resistance oscillation, in agreement with experimental findings.

\end{abstract}


\maketitle

Since the discovery of microwave induced magnetorersistance oscillations (MIMOs) 
in high-mobility two-dimensional (2D) electron gas (EG)\cite{Zud01,Ye,Mani,Zud03}
tremendous experimental\cite{Yang,Dor,Mani04,Will,Stud,Kovalev,Mani-apl,Du04,Dor04} 
and theoretical\cite{Ryz,Ryz86,Anderson,Koul,Andreev,Durst,Xie,Lei03,Dmitriev,Ryz05199,
Vav,Mikh,Dietel,Torres,Dmitriev04,Inar,Ryz0411370} interest has been attracted to 
radiation related magneto-transport in 2DEG. 
Most of previous investigations were focused on the range of low magnetic fields 
$\omega_c/\omega \leq 1$ ($\omega_c$ stands for the cyclotron frequency) 
subject to a radiation of frequency $\omega/2\pi\leq 100$\,GHz, 
where MIMOs show up strongly and Shubnikov-de Haas oscillations (SdHOs) 
are relatively weak. Recent observations clearly show that the amplitudes of SdHOs 
are also strongly affected by the microwave radiation in both the low 
($\omega_c/\omega \leq 1$) and the high ($\omega_c/\omega > 1$) 
magnetic field ranges.\cite{Kovalev,Mani-apl,Du04,Dor04}

We propose that these SdHO modulations come from the electron heating
induced by the microwave radiation. By carefully calculating the electron 
temperature based on the balance of the energy absorption from the radiation 
field and the energy dissipation to the lattice through the electron-phonon 
interactions, we reproduce all of the interesting 
phenomena of MIMOs and SdHO modulations observed in the experiments.

Consider that a dc electric field ${\bf E}_0$ and a high frequency (HF) field 
${\bf E}(t)\equiv{\bf E}_s \sin(\omega t)+{\bf E}_c\cos(\omega t)$ 
 are applied in a quasi-2D system consisting of $N_{\rm e}$ 
interacting electrons in a unit area of the $x$-$y$ plane, 
together with a magnetic field ${\bf B}=(0,0,B)$ along the $z$ direction.

For ultra-clean, high-carrier-density electron systems in the experiments 
at low temperature without the onset of the quantum Hall effect, 
the transport under a modest radiation field can be 
described by the balance-equation model in terms of 
the time-dependent electron drift velocity 
$
{\bf v}(t)={\bf v}_0-{\bf v}_1 \cos(\omega t)-{\bf v}_2 \sin(\omega t)
$, together with an electron temperature $T_{\rm e}$ characterizing
the electron heating.\cite{Lei85} They can be determined by
the following force- and energy-balance equations:\cite{Liu}
\begin{eqnarray}
{\bf v}_1 \omega \sin(\omega t)&-&{\bf v}_2 \omega \cos(\omega t)=
\frac{1}{N_{\rm e}m}{\bf F}(t)\nonumber\\ 
\hspace{0.8cm}&+&\frac{e}{m} \left[{\bf E}_0+{\bf E}(t)+
{\bf v}(t)\times {\bf B}\right],\label{eqnf}
\end{eqnarray}
\begin{equation}
N_{\rm e}{\bf E}_0\cdot {\bf v}_0+S_{\rm p}- W=0.
\label{eqne}
\end{equation}
Here $m$ is the electron effective mass, ${\bf F}(t)$ is the damping force 
of the moving electrons, 
\begin{eqnarray}
S_{\rm p}=\sum_{{\bf q}_\|}\left| U({\bf q}_\|%
)\right| ^{2}%
\sum_{n=-\infty }^{\infty }n\omega J_{n}^{2}(\xi )\Pi _{2}({\bf %
q}_\|,\omega_0-n\omega )\,\,\,\,\,&&\nonumber\\
+\sum_{{\bf q}}\left| M({\bf q})\right|
^{2}\sum_{n=-\infty
}^{\infty }n\omega J_{n}^{2}(\xi )\Lambda _{2}({\bf q},\omega_0+
\Omega _{{\bf q}}-n\omega )&&
 \label{eqsp}
\end{eqnarray}
is the time-averaged rate of the electron energy absorption from the HF field, and 
\begin{equation}
W=\sum_{{\bf q}}\left| M({\bf q})\right|
^{2}\sum_{n=-\infty
}^{\infty } \Omega_{\bf q}J_{n}^{2}(\xi )\Lambda _{2}({\bf q},\omega_0+
\Omega _{{\bf q}}-n\omega )
 \label{eqw}
\end{equation}
is the time-averaged rate of the electron energy loss to the lattice 
due to electron-phonon scatterings. 
In the above expressions, $J_n(\xi)$ is the Bessel function of order $n$,
$
\xi\equiv \sqrt{({\bf q}_\|\cdot {\bf v}_1)^2+
({\bf q}_\|\cdot {\bf v}_2)^2}/{\omega}
$;
$\omega_0\equiv {\bf q}_\|\cdot {\bf v}_0$,
$U({\bf q}_\|)$ and $M({\bf q})$ stand for effective impurity and phonon
scattering potentials, 
$\Pi_2({\bf q}_\|,\Omega)$ and
$\Lambda_2({\bf q},\Omega)=2\Pi_2({\bf q}_\|,\Omega)
[n(\Omega_{\bf q}/T)-n(\Omega/T_{\rm e})]
$ (with $n(x)\equiv 1/({\rm e}^x-1)$)
are the imaginary parts of the electron density correlation function 
and electron-phonon correlation function in the magnetic field.

Transverse and longitudinal photoresistivities are obtained directly from  
the dc part of the force-balance equation.
The linear magnetoresistivity (${\bf v}_0\rightarrow 0$) is given by 
\begin{eqnarray}
R_{xx}&=&-\sum_{{\bf q}_\|}q_x^2\frac{|
U({\bf q}_\|)| ^2}{N_{\rm e}^2 e^2}\sum_{n=-\infty }^\infty J_n^2(\xi)\left. 
\frac {\partial \Pi_2}{\partial\, \Omega }\right|_{\Omega =n\omega }\nonumber\\
&&- \sum_{ {\bf q}} q_x^2\frac{\left| M ( {\bf
q})\right| ^2}{N_{\rm e}^2 e^2}\sum_{n=-\infty }^\infty J_n^2(\xi)\left. 
\frac {\partial \Lambda_2}{\partial\, \Omega }\right|_{\Omega =
\Omega_{{\bf q}}+n\omega}.
\,\,\,\,\,\,\,\label{rxx}
\end{eqnarray}

The $\Pi_2({\bf q}_{\|}, \Omega)$ function of the 2D
system in a magnetic field can be calculated by means of Landau representation:\cite{Ting}
\begin{eqnarray}
&&\hspace{-0.7cm}\Pi _2({\bf q}_{\|},\Omega ) =  \frac 1{2\pi
l_{\rm B}^2}\sum_{n,n'}C_{n,n'}(l_{\rm B}^2q_{\|}^2/2) 
\Pi _2(n,n',\Omega),
\label{pi_2}\\
&&\hspace{-0.7cm}\Pi _2(n,n',\Omega)=-\frac2\pi \int d\varepsilon
\left [ f(\varepsilon )- f(\varepsilon +\Omega)\right ]\nonumber\\
&&\,\hspace{2cm}\times\,\,{\rm Im}G_n(\varepsilon +\Omega){\rm Im}G_{n'}(\varepsilon ),
\end{eqnarray}
where $l_{\rm B}=\sqrt{1/|eB|}$ is the magnetic length,
$
C_{n,n+l}(Y)\equiv n![(n+l)!]^{-1}Y^le^{-Y}[L_n^l(Y)]^2
$
with $L_n^l(Y)$ the associate Laguerre polynomial, $f(\varepsilon
)=\{\exp [(\varepsilon -\mu)/T_{\rm e}]+1\}^{-1}$ is the Fermi distribution
function at electron temperature $T_{\rm e}$. 
In the case of separated levels discussed in this letter, the density of states 
of the $n$-th Landau level is modeled by a semielliptic form:\cite{Ando} 
\begin{equation}
{\rm Im}G_n(\varepsilon)=
-(2/\Gamma^{2})[\Gamma^{2}-(\varepsilon-\varepsilon_n)^{2}]^{\frac{1}{2}}
\label{ellip}
\end{equation}
around the level center $\varepsilon_n$ within half-width
$
\Gamma=(8e\omega_c\alpha/\pi m \mu_0)^{1/2},
$
and ${\rm Im}G_n(\varepsilon)=0$ elsewhere. Here $\mu_0$ is the linear mobility 
at lattice temperature $T$ in the absence of magnetic field 
and $\alpha$ is a semiempirical parameter.

Assume that the 2DEG is contained in a thin sample suspended in a vacuum 
at plane $z=0$.  
When an electromagnetic wave illuminates the plane perpendicularly 
with the incident electric field 
${\bf E}_{\rm i}(t)={\bf E}_{{\rm i}s}\sin(\omega t)+ {\bf E}_{{\rm i}c}\cos(\omega t)$, 
the HF electric field in the 2DEG determined by the Maxwell equations is   
\begin{equation}
{\bf E}(t)=\frac{N_{\rm e}e\,{\bf v}(t)}{2\epsilon_0 c}+{\bf E}_{\rm i}(t).
\end{equation}
With this ${\bf E}(t)$,
${\bf v}_1$ and ${\bf v}_2$ are explicitly solved from Eq.\,(\ref{eqnf}) for clean 
systems at low temperatures.

For systems used in the experiments at temperature $T\leq 1$\,K,
the dominant contribution to the energy absorption $S_{\rm p}$ and photoresistivity 
$R_{xx}-R_{xx}(0)$ come from the impurity-assisted photon absorption and emission 
process. 
At different magnetic field strength, this process is associated with
electron transition between either inter-Landau level states or intra-Landau-level
states. The condition 
for inter-Landau level transition with impurity-assisted single-photon 
process\cite{note1} is $\omega>\omega_c-2\Gamma$, or
$\omega_c/\omega<a_{\rm inter}=(\beta+\sqrt{\beta^2+4})^2/4$; 
and that for impurity-assisted intra-Landau level transition is 
$\omega<2\Gamma$, or $\omega_c/\omega>a_{\rm intra}=\beta^{-2}$, 
here $\beta=(32e\alpha/\pi m \mu_0\omega)^{\frac{1}{2}}$. 
To obtain the energy dissipation rate $W$, which is needed for calculating
the electron heating, we take account of
scatterings from bulk longitudinal and transverse acoustic 
phonons (via the deformation 
potential and piezoelectric couplings), as well as from longitudinal optical  
phonons (via the Fr\"{o}hlich coupling) in the GaAs-based system. 
In this letter, background charged impurities are assumed to be the dominant elastic 
scatterers and all the calculations were carried out with the $x$-direction 
(parallel to ${\bf E}_0$) linearly polarized incident microwave fields  
[${\bf E}_{{\rm i}s}=(E_{{\rm i}s},0), {\bf E}_{{\rm i}c}=0$], using  
the widely accepted material and coupling parameters bulk of GaAs.\cite{Lei851} 

\begin{figure}
\includegraphics [width=0.49\textwidth,clip=on] {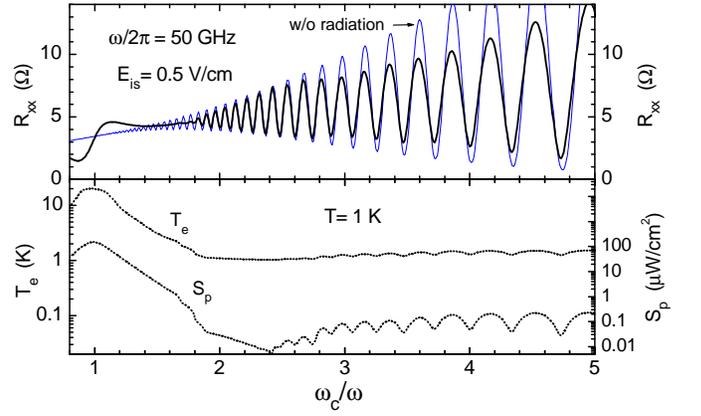}
\vspace*{-0.2cm}
\caption{The magnetoresistivity $R_{xx}$, electron 
temperature $T_{\rm e}$ and energy absorption rate $S_{\rm p}$ of a GaAs-based 2DEG 
with $N_{\rm e}=3.0\times 10^{15}$\,m$^{-2}$, $\mu_0=1000$\,m$^2$/Vs and $\alpha=5$, 
subjected to an incident HF field  
of frequency 50\,GHz with amplitude $E_{{\rm i}s}=0.5$\,V/cm at
lattice temperature $T=1 \,K$.}
\label{fig1}
\end{figure}
\begin{figure}
\includegraphics [width=0.45\textwidth,clip=on] {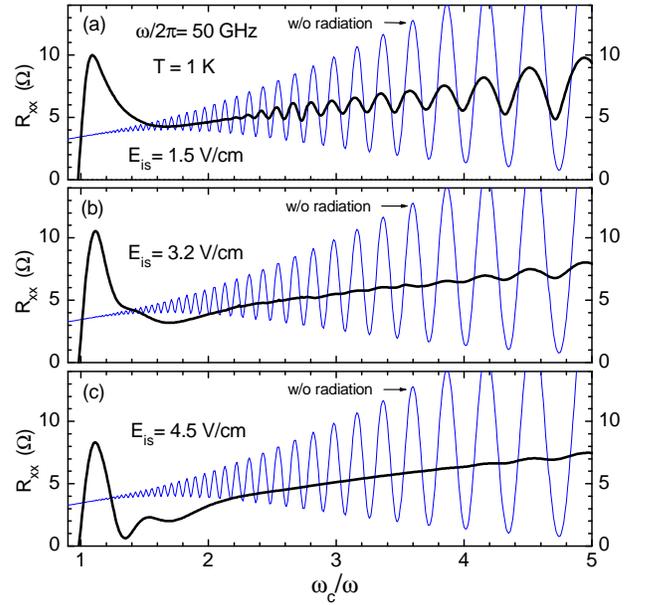}
\vspace*{-0.2cm}
\caption{Magnetoresistivity $R_{xx}$ versus $\omega_c/\omega$ 
for the same system as described in Fig.\,1, subjected to 50\,GHz 
incident HF fields $E_{{\rm i}s}\sin(\omega t)$ of three different strengths at $T=1$\,K.}
\label{fig2}
\end{figure}
\begin{figure}
\includegraphics [width=0.436\textwidth,clip=on] {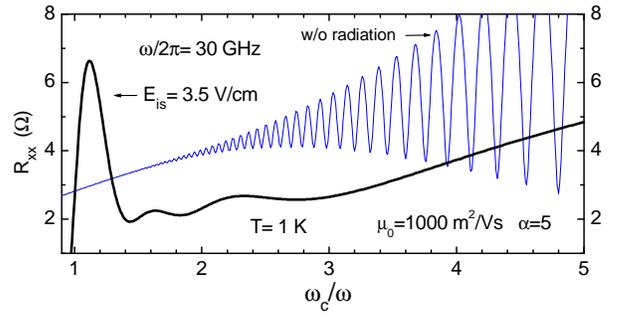}
\vspace*{-0.2cm}
\caption{Magnetoresistivity $R_{xx}$ versus $\omega_c/\omega$ 
for the same system as described in Fig.\,1, subjected to a 30\,GHz 
incident HF field of $E_{{\rm i}s}=3.5$\,V/cm at $T=1$\,K.}
\label{fig3}
\end{figure}

Figure 1 shows the calculated energy absorption rate $S_{\rm p}$, the electron 
temperature $T_{\rm e}$, and the longitudinal resistivity $R_{xx}$ 
as functions of $\omega_c/\omega$ for a 2D system having electron density 
$N_{\rm e}=3.0\times 10^{15}$\,m$^{-2}$, 
linear mobility $\mu_0=1000$\,m$^2$/Vs, and broadening parameter 
$\alpha=5$, illuminated by a microwave radiation of frequency 
$\omega/2\pi=50$\,GHz and amplitude $E_{{\rm i}s}=0.5$\,V/cm at lattice temperature 
$T=1$\,K. The energy absorption rate $S_{\rm p}$ exhibits a large main peak 
at cyclotron resonance $\omega_c/\omega=1$ 
and so does the electron temperature $T_{\rm e}$. 
For this GaAs system, $\beta=0.65$, $a_{\rm inter}=1.9$,
and $a_{\rm intra}=2.4$. We can see that, at lower magnetic fields, especially
when $\omega_c/\omega<1.5$,  the system absorbs enough energy from the radiation 
field via inter-Landau level transitions and $T_{\rm e}$ is significantly 
higher than $T$, 
with the maximum as high as 21\,K around $\omega_c/\omega=1$. 
With increasing strength of the magnetic field, the inter-Landau level transition 
weakens and the 
absorbed energy and electron temperature decreases rapidly. Since the intra-Landau level 
transitions begin to appear at $\omega_c/\omega>2.4$, there is a magnetic field range 
within which the impurity-assisted photon absorption and emission process is very weak, 
such that very little radiation energy is absorbed and the electron temperature 
$T_{\rm e}$ is almost equal to the lattice temperature $T$. The
magnetoresistivity $R_{xx}$ showing in the upper part of Fig.\,1, exhibits
interesting features.  MIMOs clearly appear at lower magnetic fields, which is  
insensitive to the electron heating even at $T_{\rm e}$ of order of 20\,K. 
SdHOs appearing in the higher 
magnetic field side, however, are damped due to the 
rise of the electron temperature $T_{\rm e}>1$\,K as compared to that without
radiation. However, in the range of $1.85<\omega_c/\omega<2.4$, where electrons
are essentially not heated, the SdHOs are almost not affected by the 
microwave. When the microwave amplitude increases to $E_{{\rm i}s}=1.5$\,V/cm 
(Fig.\,2a), MIMOs become strong and SdHOs greatly damped. 
At $E_{{\rm i}s}=3.2$\,V/cm (Fig.\,2b), SdHOs almost disappear and both real 
and virtual multi-photon processes show up in the MIMOs, resulting in a 
descent of the magnetoresistivity $R_{xx}$ down below the average value of
its oscillatory curve without radiation. This magnetoresistivity descent  
becomes quite strong at $E_{{\rm i}s}=4.5$\,V/cm due to enhanced multi-photon
processes as shown in Fig.\,2c.

\begin{figure}
\includegraphics [width=0.47\textwidth,clip=on] {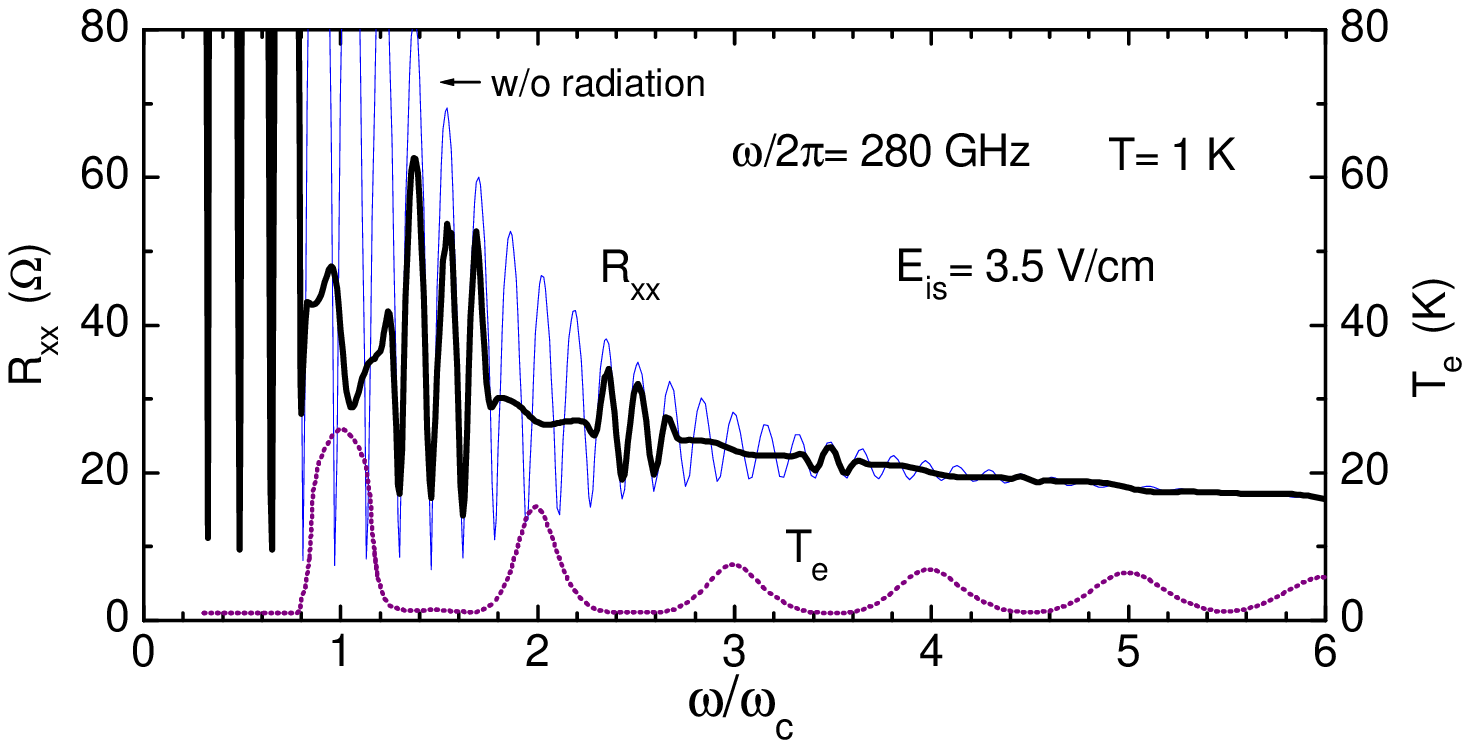}
\vspace*{-0.2cm}
\caption{The magnetoresistivity $R_{xx}$ of a GaAs-based 2DEG with 
$N_{\rm e}=2.0\times 10^{15}$\,m$^{-2}$, $\mu_0=500$\,m$^2$/Vs and $\alpha=1$, 
subjected to a 280\,GHz HF field of $E_{{\rm i}s}=3.5$\,V/cm at $T=1$\,K.}
\label{fig3}
\end{figure}

The radiation-induced $R_{xx}$ suppression at $\omega_c/\omega >1 $  
appears even more remarkable with lower frequency irradiation. Figure 3 shows 
$R_{xx}$-versus-$\omega_c/\omega$ for the same system as described in Fig.\,1, 
subjected to a 30\,GHz microwave of strength $E_{{\rm i}s}=3.5$\,V/cm at $T=1$\,K. 
At this frequency, the ranges for 
intra-Landau level and inter-Landau level single-photon transitions 
overlap. The enhanced effect of virtual and real multi-photon processes pushes
the resistivity $R_{xx}$ down below its zero-radiation oscillatory curve 
across a wide range of $\omega_c/\omega >1$, 
in agreement with experimental observations.\cite{Mani-apl,Dor04}
The microwave-induced suppression of the dissipative magnetoresistivity may be 
further enhanced by the effect of dynamic localization.\cite{Ryz0411370}

The radiation modulation of SdHOs can be seen at lower magnetic field range 
$\omega_c/\omega<1$ under higher frequency microwave illumination with 
simultaneously appearing of MIMOs.
Figure 4 shows the calculated $R_{xx}$ and $T_{\rm e}$ 
for a 2D system of electron density 
$N_{\rm e}=2.0\times 10^{15}$\,m$^{-2}$, linear mobility 
$\mu_0=500$\,m$^2$/Vs and $\alpha=1$, subject to a 280\,GHz microwave radiation 
of amplitude $E_{{\rm i}s}= 3.5$\,V/cm at lattice temperature $T=1$\,K. One can clearly 
see peaks of the electron temperature $T_{\rm e}$ and nodes of SdHO modulation
at $\omega/\omega_c=1,2,3,4$ and 5, together with MIMOs. 
These are in agreement with the experimental observation reported in 
Ref.\onlinecite{Kovalev}. 

This work was supported by Projects of the National Science Foundation of China
 (60676041 and 10390162), the Special Funds for Major State Basic Research Project, 
 and the Shanghai Municipal Commission of Science and Technology.
 
%\newpage 


\begin{thebibliography}{99}
\bibitem{note1} Mulitiphoton ($|n|>1$) processes can take place up to higher magnetic 
fields, but they are usually much weaker than the single-photon ($n =\pm 1$) process.
\end{thebibliography}
\end{document}